# Charge to Spin Conversion in van der Waals Metal NbSe$_2$

Anamul Md. Hoque[1], Bing Zhao[1], Dmitrii Khokhriakov[1], Prasanta Muduli[2], Saroj P. Dash[1,3*]

[1]Department of Microtechnology and Nanoscience, Chalmers University of Technology, SE-41296, Göteborg, Sweden

[2]Department of Physics, Indian Institute of Technology Madras, Chennai 600036, India

[3]Graphene Center, Chalmers University of Technology, SE-41296, Göteborg, Sweden

**Abstract:**

Quantum materials with a large charge current-induced spin polarization are promising for next-generation all-electrical spintronic science and technology. Van der Waals metals with high spin-orbit coupling and novel spin textures have attracted significant attention for an efficient charge to spin conversion process. Here, we demonstrate electrical generation of spin polarization in NbSe$_2$ up to room temperature. To probe the current-induced spin polarization in NbSe$_2$, we used a graphene-based non-local spin-valve device, where the spin-polarization in NbSe$_2$ is efficiently injected and detected using non-local spin-switch and Hanle spin precession measurements. A significantly higher charge-spin conversion in NbSe$_2$ is observed at a lower temperature, below the superconducting transition temperature $T_c \sim 7$ K of NbSe$_2$. However, the charge-spin conversion signal could only be observed with a higher bias current above the superconducting critical current, limiting the observation of signal only to the non-superconducting state of NbSe$_2$. Systematic measurements provide the possible origins of the spin polarization to be predominantly due to the spin Hall effect or Rashba-Edelstein effect in NbSe$_2$, considering different symmetry allowed charge-spin conversion processes.

## Introduction

Two-dimensional (2D) materials and their van der Waals (vdW) heterostructures have become attractive platforms to explore numerous physical phenomena, primarily associated with spin-orbit coupling (SOC), exotic superconductivity, and magnetism for next-generation novel electronic devices[1–3]. 2D transition metal dichalcogenides (TMDCs) are ideal hosts for realizing spin-polarized electronic states due to the high SOC accompanied with broken symmetries (structural, rotational, translational, etc.) in the crystal structure[4–7]. Recently, van der Waals materials have paved the way for utilizing various charge-to-spin interconversion (CSC) processes owing to fascinating electronic band structures and lower crystal symmetries[4,8–10]. For example, both conventional and unconventional charge-to-spin interconversion effects and out-of-plane spin-orbit torque (SOT) components are observed in semimetals such as $WTe_2$, $MoTe_2$, and $NbSe_2$ up to room temperature[11–18]. Interestingly, the 2D materials can be fabricated in vdW heterostructures with atomically clean interfaces without adulterating their distinctive electronic properties, which provide new routes for band structure engineering and proximity induced SOC[19–22]. For instance, heterostructures of graphene with 2D semiconductors, semimetals, and topological insulators have shown unprecedented gate tunable CSC processes and novel spin textures[17,23–31].

It is known that 2H $NbSe_2$ is a vdW layered metallic TMDC with superconducting (SC) behaviour below a critical temperature $T_c \approx 7K$[32–34]. It is expected that the CSC effects can be enhanced in the SC state mediated by quasiparticles with higher spin lifetimes[3,35,36]. Interestingly, $NbSe_2$ in the normal state also has enormous prospects in the CSC process, triggered by high SOC of the Nb $4d$ orbital and breaking of symmetries with higher electrical conductivities than semiconductive and semimetallic TMDCs[11]. $NbSe_2$ also exhibits Ising-type SOC similar to the intrinsic Zeeman field, which results in unconventional spin textures[37–39]. Recently in the $NbSe_2$ and Permalloy bilayer system, current-induced spin-orbit torque (SOT) measurements demonstrated a large anti-damping torque owing to strain-induced symmetry breaking[11]. Furthermore, spin- and angle-resolved photoemission spectroscopy (ARPES) also reveals that the electronic band structure of $NbSe_2$ in the normal state hosts a strong momentum-dependent spin polarization at Fermi level[40,41]. Such fascinating spin-dependent electronic properties in $NbSe_2$ are highly desirable to create current-induced spin polarization.

Here, we demonstrate the electronic generation of spin polarization in $NbSe_2$ up to room temperature owing to the efficient charge to spin conversion (CSC) process. The engendered spin polarization in $NbSe_2$ is efficiently injected into the graphene spin-channel in the vdW heterostructure spin-valve device and detected by a ferromagnet (FM) using non-local spin-switch and Hanle spin-precession measurements up to room temperature. A significantly higher CSC signal in $NbSe_2$ is observed at a lower temperature, however, in the non-superconducting regime. These findings demonstrate $NbSe_2$ to be a metallic spin source up to room temperature with efficient control of spin polarization by the bias current magnitude, which can be pivotal for future energy-efficient all-electric spin-based technologies[9,10,42,43].

## Results

### Fabrication of $NbSe_2$-graphene heterostructure device

To detect the CSC properties in $NbSe_2$, we fabricated $NbSe_2$-graphene vdW heterostructure devices. We used chemical vapor deposited (CVD) graphene as a spin channel material, due to its excellent spin transport property arising from low SOC and hyperfine interactions[44–46]. Most interestingly, it was demonstrated that graphene can make a very good vdW heterostructure with other 2D materials[20,47,48]. Figure 1a presents the schematic of the $NbSe_2$-graphene vdW heterostructure device along with ferromagnetic (FM) contacts to characterize the spin transport properties in the heterostructure (see the device fabrication part for details). A scanning electron microscopic (SEM) image of a fabricated device consisting of CVD graphene, multilayer NbSe$_2$ flake, and multiple FM contacts is shown in Fig.1(b). An optical micrograph and atomic force microscopic (AFM) image of the corresponding device has been shown in Fig. S1. This device consists of 23 nm NbSe$_2$ on top of monolayer CVD graphene with TiO$_2$/Co as ferromagnetic tunnel contacts. The interface resistance between NbSe$_2$-graphene is found to be 50 Ω, the FM contact resistance is 12kΩ and the field-effect mobility (μ) of the graphene channel is ≈ 2000cm$^2$V$^{-1}$s$^{-1}$ (see Fig. S2).

As a high SOC material, NbSe$_2$ can give rise to current-induced in-plane (y-axis) spin polarization via conventional spin Hall effect (shown in Fig. 1c), where charge current ($I_c$) along the x-axis creates a transverse spin current ($I_s$) along the z-axis[15,18]. Furthermore, high SOC in conjunction with breaking inversion symmetry along the z-axis due to two different atoms in layered NbSe$_2$ crystal can result in Rashba spin splitting in the band structure with helical spin texture with opposite spin-subbands, as shown in Fig. 1(d). Upon application of an electric field (E), hence the charge current can shift the helical Fermi surface in the k-space and create net spin polarization via the Rashba-Edelstein effect. The created spin polarization in NbSe$_2$ can be injected into a graphene spin-channel in VdW heterostructure and detected by non-local measurement geometry to realize a pure spin signal.

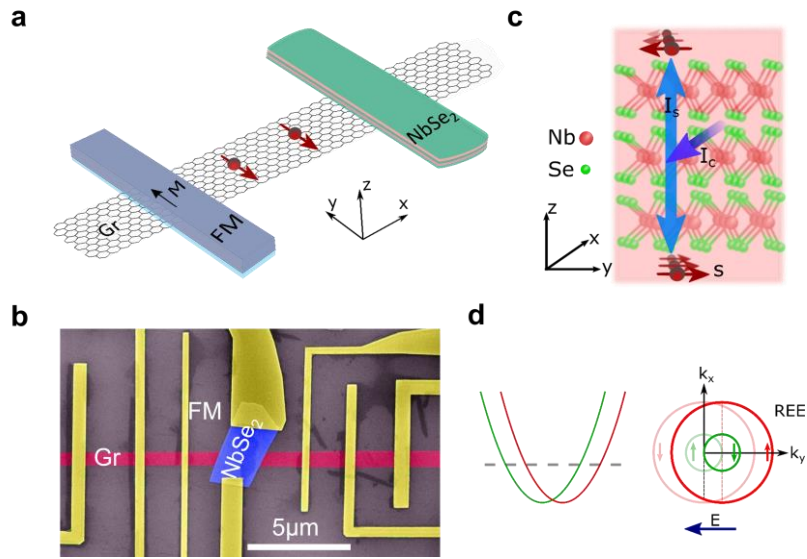

**Figure 1. The device design of NbSe$_2$-graphene heterostructure**. **(a)** Schematic of NbSe$_2$-graphene heterostructure, where NbSe$_2$ creates non-equilibrium spin polarization and injects spin-polarized electrons into graphene spin-channel, which is detected by an FM (TiO$_2$/Co) contacts in non-local measurement geometry. **(b)** Colored scanning electron microscopic (SEM) image of a fabricated device consisting of CVD graphene (Gr), multilayer NbSe$_2$ flake (blue), and multiple FM contacts (yellow) to characterize spin transport properties in the heterostructure. The scale bar (white line) in the image is 5 μm. **(c)** Schematic illustration of charge to spin conversion process due to conventional spin Hall effect in NbSe$_2$, where charge current ($I_c$) engenders a transverse spin current ($I_s$). **(d)** Charge current from the applied electric field (E) induced spin polarization due to the Rashba-Edelstein effect (REE) at the Fermi surface in spin-split bands of a high spin-orbit coupling system.

## Material properties and superconductivity of NbSe$_2$

At first, high-quality crystal structures of the materials were ensured by Raman spectroscopy. The Raman spectrum of single-layer CVD graphene (top panel) and exfoliated multilayer NbSe$_2$ (bottom panel) using 638 nm LASER have been presented in Fig. 2a. The Raman spectra of graphene confirm high-quality graphene crystal since almost no defect-induced D peak is observed. Furthermore, the higher intensity of the 2D peak at 2645 cm$^{-1}$ than that of the G peak at 1590 cm$^{-1}$ indicates the growth of single-layer graphene[49]. In the case of NbSe$_2$, the characteristic out-of-plane phonon mode A1g peak for multilayer 2H NbSe$_2$ is observed at 230 cm$^{-1}$ [50,51] (see Fig. 2a bottom panel).

We started with investigating the superconducting properties of exfoliated multilayer NbSe$_2$ flake by measuring the temperature dependence of longitudinal resistance in four-terminal measurement (4T) geometry, as shown in Fig. 2b. The device picture and measurement geometry are shown in the insets of Fig. 2b. We used 1μA of DC bias current and the superconducting critical temperature (T$_c$) in this flake is found to be 6.8K. The four-probe IV measurement with DC bias current and corresponding differential resistance (dV/dI) as a function of the bias current at 3K for the NbSe$_2$ flake are shown in Fig. 2c and 2d, respectively. The nonlinear IV is due to superconductive properties in NbSe$_2$ and from the dV/dI plot, we can estimate the critical current density, $J_{c,\ NbSe2}$ = 0.7x10$^6$ A/cm$^2$ for the multilayer NbSe$_2$ at 3K, consistent with the recent study[52]. The room temperature IV measurement (shown in Fig. S3b, c) is in agreement with the metallic properties of NbSe$_2$. Raman spectroscopies, together with the presence of superconducting transition in NbSe$_2$ confirm good-quality materials are used in the heterostructure.

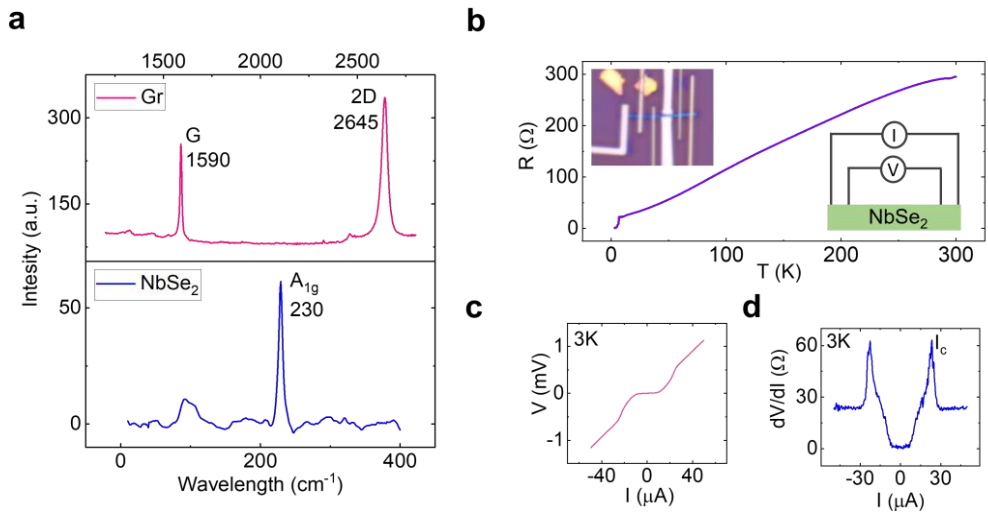

**Figure 2. Materials characteristics of NbSe$_2$ and graphene**. **(a)** Raman spectrum of single-layer CVD graphene (top panel) and exfoliated multilayer NbSe$_2$ (bottom panel) using 638 nm LASER. **(b)** Temperature dependence of the longitudinal resistance of multilayer NbSe$_2$ flake, where superconducting critical temperature (T$_c$) is around 6.8K is evident. Insets are the device picture (top) and four-terminal measurement geometry. **(c,d)** The four-probe IV measurements and corresponding differential resistance as a function of the bias current in the multilayer NbSe$_2$ flake at 3K.

**Charge-spin conversion in NbSe$_2$**

In order to investigate the charge-spin conversion (CSC) effect in NbSe$_2$, a charge current is applied in a multilayer NbSe$_2$ flake to create a non-equilibrium spin polarization on its surface. The spins are injected into the graphene channel and finally detected by an FM contact in non-local measurement geometry. Figure 3a shows a schematic illustration of the measurement geometry used to detect the CSC effect in NbSe$_2$ along with the axis orientation and corresponding spin (s), charge current (I$_c$) and spin components (I$_s$). According to our measurement geometry in Fig. 3a by considering conventional SHE in NbSe$_2$, a charge current along x-axis can create an out-of-plane (z-axis) spin current, which renders spin polarization along y-axis. It is worth mentioning that the spin diffusion current in graphene spin channel is along x-axis in our measurement geometry with spin orientation towards y-axis after spin polarization is created in NbSe$_2$ and injected into graphene. In our experiments, we measure the non-local voltage at the FM contact (left nearest contact to the NbSe$_2$ flake, shown in Fig. 1b) by applying varying magnetic fields along with the B$_y$ and B$_z$ directions, respectively. A varying magnetic field along the magnetic easy axis (y-axis) switches the magnetization of the FM contact and renders a switching signal which detects non-equilibrium in-plane spin in graphene injected from the NbSe$_2$ flake. The spin-switch signal presented in Fig. 3b is measured with I = 400 μA at V$_g$ = -40 V (the up and down magnetic sweep directions are indicated by arrows). The amplitude of the signal estimated from the change in non-local resistance corresponding to the opposite magnetization of the FM contact is about, $\Delta R_{nl}$ = 1.77 ± 0.6 mΩ.

Furthermore, an out-of-plane varying magnetic field ($B_z$) in our measurement geometry should render a Hanle spin precession signal, which unequivocally confirms the CSC process in the NbSe$_2$ and spin transport in the graphene spin channel. Figure 3(c) shows the manifested Hanle spin signal measured with I = +420 µA and $V_g$ = -40 V while injecting spin from NbSe$_2$ into the graphene channel along with the fitting to equation S1. We have estimated the spin lifetime, $\tau_s$ = 23 ± 6 ps and spin diffusion length, $\lambda_s$ = $\sqrt{\tau_s D_s}$ = 0.65 ± 0.05 µm, considering the channel length L = 2.4 µm (distance between the center of the NbSe$_2$ flake to the center of the detector's FM electrode). Moreover, the spin transport in the pristine CVD graphene at different gate voltages is shown in S4, where spin lifetime is estimated to be around 150 ps. In NbSe$_2$-graphene heterostructure, the lower $\tau_s$ in graphene spin channel after spin is injected from NbSe$_2$ can be attributed to the influence of long-range disorders, lattice deformation, and extrinsic interstitials in the graphene crystal that acts as spin-defect centers. These imperfections might be introduced during processing CVD graphene, NbSe$_2$ transfer, and device fabrication processes[53]. Additionally, spin absorption by NbSe$_2$ can also give rise to lower $\tau_s$ because of the transparent NbSe$_2$-graphene interface. Interestingly, we observe only symmetric Hanle component, although the NbSe$_2$ flake in our device is at an angle to the graphene spin channel[54]. Furthermore, any contribution from the spin injection from the FM contact on NbSe$_2$ can be eliminated from the spin switch signal, because the influence of and FM magnetization would have manifested typical spin valve signal (as presented in Fig. S4a) with double spin-valve switching while sweeping $B_y$. Besides, Hanle measurement rules out any effect of stray fields from the detector FM contact on the manifested CSC signals, since the stray Hall effect would have rendered linear Hall signal for an out-of-plane field ($B_\perp$)[55].

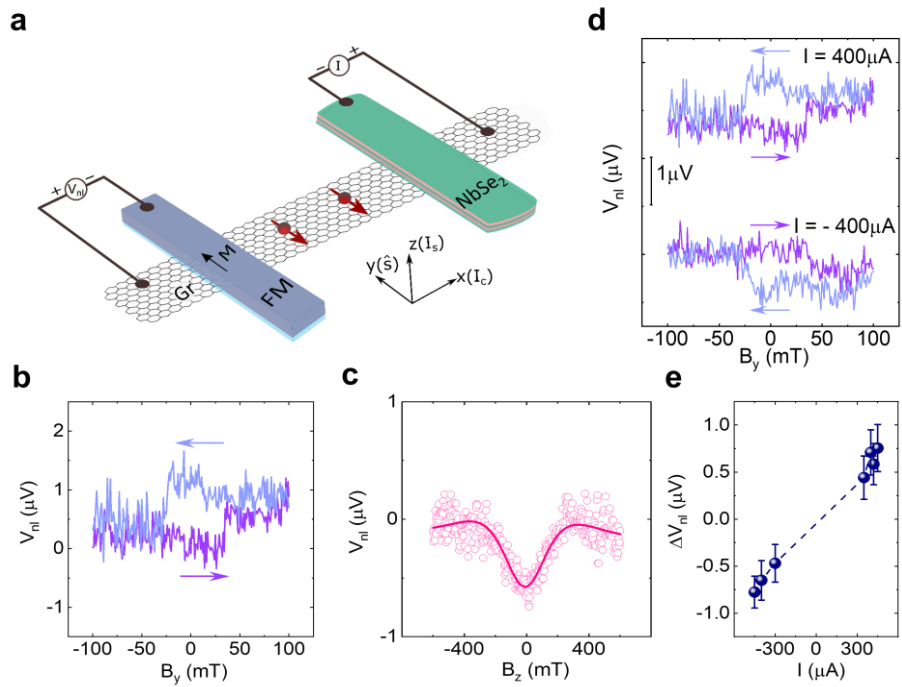

**Figure 3. Charge-spin conversion in NbSe$_2$ at room temperature. (a)** Schematic illustration of non-local (NL) measurement geometry to detect the charge-spin conversion effect in NbSe$_2$ by injecting spin current into the graphene spin channel. **(b,c)** The spin-switch and Hanle spin precession measurements for spin injection from NbSe$_2$ with a $B_y$ and $B_z$ sweep, respectively. For spin-switch experiments, the up and down magnetic sweep directions are indicated by arrows. The Hanle data is fitted using equation S1. A linear background is subtracted from the data. **(d)** Non-local spin-switch signals ($V_{nl}$) at room temperature with bias currents of I = ±400 µA and $V_g$ = -40 V. A shift is added in the y-axis for clarity. **(e)** The magnitude of the spin-switch signal with applied bias

current magnitude in NbSe$_2$-graphene heterostructure. The error bars are calculated from the noise level of the non-local signal. All the measurements were conducted in Dev 1 at room temperature.

Next, we estimated the spin polarization of NbSe$_2$ is about P$_{NbSe2}$ = 1± 0.3%, assuming the spin polarization of FM contact on bare CVD graphene is P$_{Co}$ = 1.5± 0.6% (see Supplementary Note 1)[56]. The efficiency of the CSC process due to spin Hall effect in NbSe$_2$ can be characterized by the spin Hall angle (θ$_{SH}$ ∝ J$_s$/J$_c$) and by using a simple model (as discussed in Supplementary Note 1), we found the θ$_{SH}$ of NbSe$_2$ varies approximately from 0.68 ± 0.15 to 0.30 ± 0.06 by assuming the variation of spin diffusion length in NbSe$_2$, $\lambda_{NbSe2}$ = 5 nm to 40 nm (see Fig. S5a and Supplementary Note 2)[18,57–59]. This estimation of θ$_{SH}$ is consistent with the recently reported theoretical study, where θ$_{SH}$ in NbSe$_2$ is predicted to be ≈ 0.5 by light irradiation[60]. Note that the spin diffusion length ($\lambda_{NbSe2}$) in NbSe$_2$ is not experimentally reported yet. We also analytically calculated the length scale, θ$_{SH}$.$\lambda_{NbSe2}$ (nm), associated with the CSC in NbSe$_2$ (5-13nm) (see Fig. S5b), which is comparable to the recently reported length scale in layered TMDCs, e.g., 5nm in WSe$_2$, 1.15nm in MoTe$_2$[15,31]. Another plausible origin of the measured CSC signal could be the Rashba-Edelstein effect (REE) due to Rashba spin-split bands in NbSe$_2$ and its proximity effect in the NbSe$_2$-graphene heterostructure region[5]. The characteristic efficiency parameter ($\alpha_{RE}$) of the REE is calculated (Supplementary Note 2) to be 5.3 ± 1.8%, which is consistent with the recent studies on 2D material heterostructures[15,24,25,29,30].

We systematically measured the electrical bias dependence of the CSC signal in NbSe$_2$, as shown in Fig. 3(d) for the spin-switch signal (V$_{nl}$). It is evident that reversing the bias current direction results in opposite spin-switch signals due to the reversal of the accumulated spin polarization direction. A full bias-dependent spin-switch measurement was carried out and the amplitudes of the spin-switch signal (ΔV$_{nl}$) at different bias current magnitudes are summarized in Fig. 3e with a linear fitting (dashed line). It can be seen that with increasing bias current the ΔV$_{nl}$ increases linearly because a larger bias current generates more spin polarization in NbSe$_2$, which is eventually injected and measured in the graphene spin channel.

## Gate dependence of charge to spin conversion signals in NbSe$_2$

To verify the influence of gate voltage on the CSC singals in NbSe$_2$ and spin injection into graphene channel, we performed backgate (V$_g$) dependent measurement of the CSC signals in our devices with the same measurement scheme presented in Fig. 3a. Figure 4a shows the measured non-local spin-switch signals as a function of varying in-plane magnetic field (B$_y$) at various V$_g$ = - 40 V to 50 V for both n and p doped graphene regime at room temperature. The CSC signals in NbSe$_2$ and spin injection into graphene are only observed at the higher negative gate voltages. The Hanle spin-precession signals of the CSC effect in our NbSe$_2$-graphene heterostructure with changing out-of-plane magnetic field (B$_z$) at different gate voltages are shown in Fig. 4b along with the fitting to equation S1 (solid line). It can be seen that the signals are only also present at the higher negative V$_g$ like spin switch signal.

Figure 4c shows the amplitude of the spin-switch signal, ΔV$_{nl}$ (top panel), and estimated spin lifetime (middle panel), τ$_s$, of the spin transport in graphene channel after spin is injected from NbSe$_2$ by CSC process as a function of V$_g$. To compare the spin transport properties in the graphene channel after the spin is injected from NbSe$_2$, the NbSe$_2$-graphene heterostructure channel resistance at different gate voltages is depicted in the bottom panel of Fig. 4c. The Dirac point of the NbSe$_2$-graphene heterostructure region is about 33V. It can be seen that, when the graphene channel resistance increases for V$_g$ > -10V , the spin injection efficiency decreases due to conductivity mismatch issue between NbSe$_2$ and graphene[61]. The spin transport singal in pristine CVD is shown in Fig. S4c for all the gate voltages in n- and p-doped regimes with a modulation near Dirac point,which can be attributed to the conductivity mismatch metween the graphene and the FM contact. The absence of the spin signals, injected into graphene channel from NbSe$_2$, at the positive gate voltages can also be due to a strong spin absorption by NbSe$_2$[26].

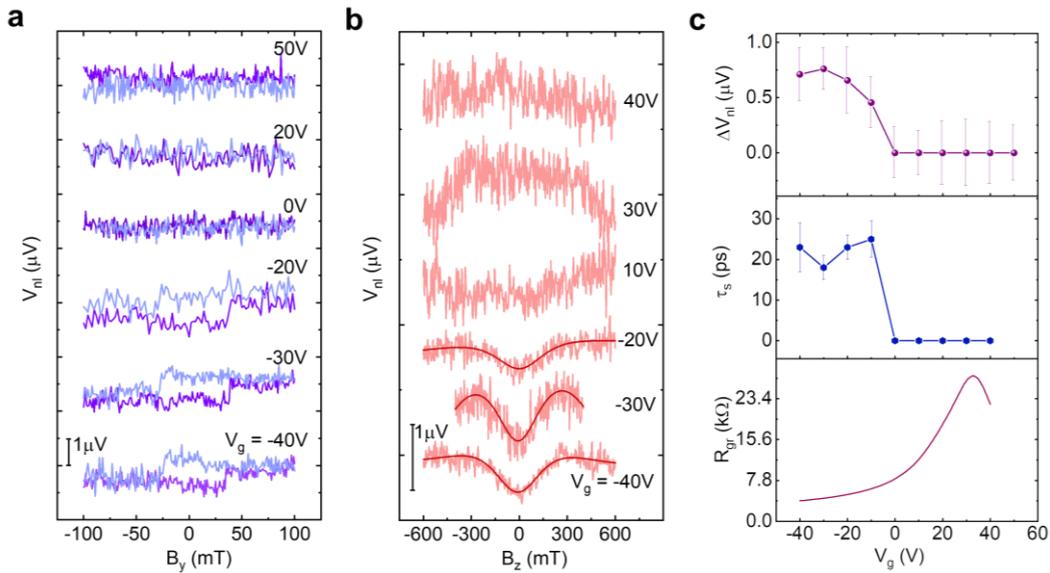

**Figure 4: Gate dependence of charge to spin conversion signals in NbSe$_2$, and spin injection into graphene spin channel**. **(a)** The non-local spin-switch signals as a function of in-plane field (B$_y$) at various V$_g$ = - 40 V to 50 V for both n and p doped graphene regime with a constant bias current I = 400 µA at room temperature. A linear background has been subtracted from the measured data and y-axis shift is added for clear depiction. **(b)** The evolution of non-local Hanle spin-precession signals as a function of out-of-plane field (B$_z$) at various V$_g$ = - 40 V to 40 V with I = 420 µA for both n and p doped graphene regime with fitting to S1 (solid line). **(c)** The magnitude of the spin-switch signal ΔV$_{nl}$ as a function of V$_g$ (top panel). Error bars are estimated from the noise level of the signal. Back-gate dependence of the spin lifetime ($\tau_s$) of the injected spin in graphene from NbSe$_2$ (middle panel). Error bars are estimated from the data fitting. The NbSe$_2$-graphene heterostructure channel resistance at different gate voltages (bottom panel). The measurements were performed in Dev 1 at room temperature.

## Temperature dependence of charge-spin conversion in NbSe$_2$

We measured the temperature dependence of the CSC process in NbSe$_2$ to observe the evolution of spin polarization in NbSe$_2$ with temperature in Dev2 (the device picture is shown in Fig. S1d). Figure 5a shows the non-local spin-switch signal arising due to the CSC effect in NbSe$_2$ and subsequent spin injection into graphene at 3K. The magnitude of the CSC signal is found to be $\Delta R_{nl} \approx 106 \pm 27$ mΩ. Next, we measured the Hanle spin precession signal above and below T$_c$ of NbSe$_2$ to validate the manifested spin-switch signal is a spin-related phenomenon. Fig. 5b shows the Hanle signals along with the fitting to equation S1. The magnitude of the Hanle spin signal, ΔR$_{nl}$, and extracted spin lifetime $\tau_s$ at different temperatures (3k to 30K) have been shown in Fig. 5c. Interestingly we found that the ΔR$_{nl}$ increases drastically below T$_c$, but $\tau_s$ remains unchanged around 150 ps, below and above T$_c$, because spin transport parameters of graphene are known to be weakly dependent on temperature[14,45]. We would like to note that, as $\tau_s$ remains unchanged, the larger ΔR$_{nl}$ can be attributed to more efficient CSC conversion effect below T$_c$. However, the increase in CSC signal with decreasing temperature could also be due to the decrease of NbSe$_2$ resistivity ($\rho_{NbSe2}$) and conductivity mismatch between NbSe$_2$ and graphene interface.

It is expected that quasiparticle mediated CSC process to be enhanced near the superconducting state of the corresponding material[35,62]. Besides, the spin lifetime can be extremely high in the SC state of NbSe$_2$ because it takes a longer time for the quasiparticles to scatter by the spin-orbit impurities in comparison to the normal electron scattering rates[3,35,36]. Although we manifested CSC signal in NbSe$_2$ at 3K (below Tc~7K), but we would also like to mention that the NbSe$_2$ flake is not in the superconducting state in our CSC experiments as the applied bias current density (~4x10$^6$ A/cm$^2$) is much higher than the critical current density (~0.7x10$^6$ A/cm$^2$ at 3K). We optimized the bias current magnitude to maximize the CSC signal at different temperatures that surmount the noise level of the signals, which vary due to the conductivity mismatch of the NbSe$_2$ and detector FM contact with the graphene spin channel.

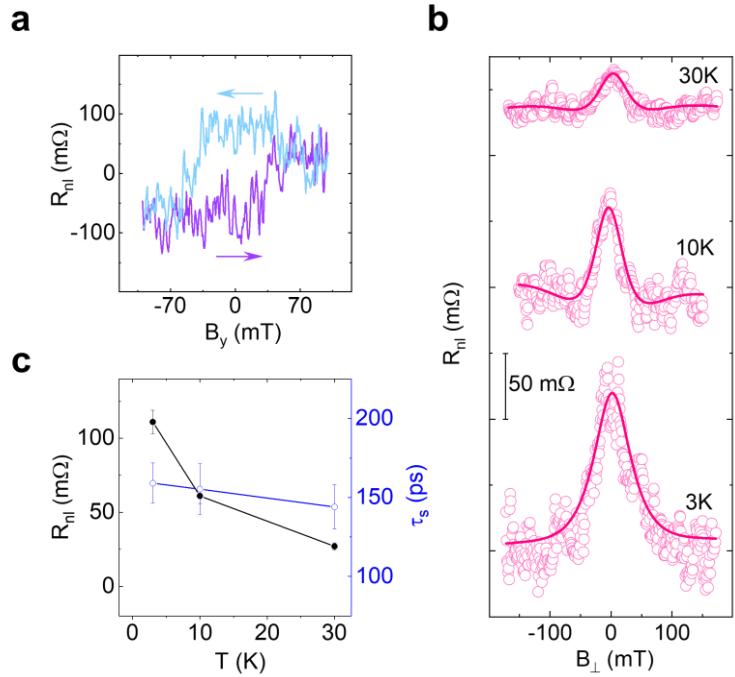

**Figure 5. Temperature dependence of charge-spin conversion effect in NbSe$_2$.** (a) Spin-switch signal for charge-spin conversion effect in NbSe$_2$ (Dev. 2) resulting in spin injection from NbSe$_2$ into graphene at 3K. (b) Hanle spin precession measurements along with the fitting to equation S1 (solid line). Measurements were performed in Dev 2 in a bias current range of 200 - 500 µA and at $V_g$ = -40 V. A linear background is subtracted from the data and shifted along the y-axis for clear visualization. (c) The magnitude of the Hanle spin signal ΔR$_{nl}$ and extracted spin lifetime ($\tau_s$) at different temperatures.

## Discussion

Here, we discuss possible origins that can give rise to the manifested CSC signal in NbSe$_2$. First and foremost, the non-local spin-switch with varying in-plane B$_y$ field and Hanle spin-precession measurements with changing out-of-plane field B$_z$ confirm that the detected signals are due to the in-plane (y-axis) spin-polarized current that is created in NbSe$_2$ and injected into graphene[56,63]. Furthermore, the manifestation of linear bias dependence and a sign reversal behavior with opposite bias current directions rule out the thermal contributions in the measured CSC signal, as the thermal effects should not change with the bias direction and deviate the linear bias dependence[24,25]. Finally, we can also rule out the spin polarization generation via proximity induced spin Hall effect (SHE) because this effect would have resulted in an out-of-plane spin polarization (s$_z$) in the heterostructure region, which is not measured[64]. We can also discard the possible impact of the Ising states on the measured CSC signal since Ising spins in the SC state would have turned spin polarization into out of plane (z-axis) direction in NbSe$_2$, which is not manifested, and our CSC measurements are limited to the non-SC state in NbSe$_2$. In addition, proximity induced Rashba-Edelstein (REE) effect in graphene from NbSe$_2$ should have rendered opposite sign of the measured CSC signal for the p- and n-doped regimes[23,25,30]. However, we observed CSC signal only in the p-doped regime of graphene with higher negative gate voltage, most likely due to the conductivity mismatch issues of the graphene channel with NbSe$_2$ and FM contacts in the n-doped regime of graphene (see Fig. S2a)[26,28,61]. Hence, proximity induced REE cannot be ruled out or nor be claimed to be the origin of the observed CSC signal with our measuremts. Furthermore, unconventional CSC in NbSe$_2$ cannot also be disregarded also since NbSe$_2$ is a layered material with broken inversion symmetry in the crystal structure, which can be further enhanced by the induced strain at the vdW heterostructure device geometry[14,15]. Finally, considering the symmetry principle[15], spin polarization direction (s) is set perpendicular to the applied charge current ($I_c$) and spin current ($I_s$) direction; the SHE and REE in $NbSe_2$ most likely merge or independently produce the observed CSC signal in $NbSe_2$. In future, the measurements of the CSC and inverse CSC in $NbSe_2$ with different device configurations and measurement geometries with different thicknesses and their correlation with properties in the superconducting state can be interesting.

## Summary

We demonstrated CSC in the normal metallic state of $NbSe_2$ up to room temperature via the spin Hall effect and Rashba Edelstein effect. . The engendered spin polarization can be injected into the graphene channel and detected in non-local measurement geometry via spin-switch and Hanle spin precession measurements. The observed gate electric field dependent modulation of the CSC signal in the graphene-$NbSe_2$ spin-valve device can be due to conductivity mismatch issues or spin absorption processes. Moreover, a higher CSC signal in $NbSe_2$ is detected at a lower temperature, however, in its non-superconducting state because of the requirement of a higher bias current than the critical current of $NbSe_2$ for observation of spin signals. Systematic measurements of the spin-switch and Hanle signals reveal that the possible origins of the in-plane spin polarization are predominantly due to the spin Hall effect or Rashba-Edelstein effect in $NbSe_2$ considering different symmetry-permitted CSC processes. Such features of current-induced spin polarization in $NbSe_2$ have promising potentials to be used as a non-magnetic spin source in future all-electric spintronic devices and spin-orbit technologies. Furthermore, the realization of CSC in superconducting quantum materials with high SOC strength can enhance the spintronic device performance by generating a larger spin current with a longer spin lifetime[3,36,65].

## Methods

Device fabrication: To fabricate $NbSe_2$-graphene heterostructure devices, first, CVD graphene (from Grolltex Inc) stripes were prepared on $SiO_2$(285 nm)/n-doped Si substrate by electron beam lithography and oxygen plasma etching. $NbSe_2$ flakes were exfoliated from bulk crystal (from Hq Graphene) by scotch tape and directly transferred onto the CVD graphene channel inside a glovebox in $N_2$ gas environment. Contacts to graphene and $NbSe_2$ were defined by electron beam lithography, electron beam evaporation, and the lift-off process. For the preparation of tunnel contacts, first, a total ~1 nm of Ti was evaporated at a base pressure less than $3\times10^{-7}$ Torr and in-situ oxidized for 30 minutes at above 30 Torr. Subsequently, 90 nm of Co was deposited in the same deposition chamber without breaking the vacuum. Electrical Measurements: The charge and spin transport measurements are performed in the magneto-transport measurement system and Physical Property Measurement System using a Keithley 6221 current source, a Keithley 2182A nano voltmeter, and Keithley 2612B source meter was used to apply gate voltages.

## Acknowledgments

The authors acknowledge financial support from EU Graphene Flagship (Core 3, No. 881603), Swedish Research Council VR project grants (No. 2021–04821), 2D TECH VINNOVA competence center (No. 2019-00068), FLAG-ERA project 2DSOTECH (VR No. 2021-05925), Graphene center, EI Nano, and AoA Materials program at Chalmers University of Technology. We acknowledge the help of staff at Quantum Device Physics and Nanofabrication laboratory in our MC2 department at Chalmers. Devices were fabricated at the Nanofabrication Laboratory, Myfab, MC2, Chalmers.